\documentclass[amssymb, preprintnumbers, showpacs, showkeys,aps,prb,superscriptaddress,twocolumn,nobibnotes]{revtex4-1}

\usepackage{pgfplots}
\usepackage{tikz}
\usetikzlibrary{positioning}

\usepackage{color} 
\usepackage{xcolor} 
\usepackage{hyperref}
\usepackage{slashed}
\usepackage{graphicx}
\usepackage{amsmath}
\usepackage{latexsym}
\usepackage{epstopdf}
\usepackage{amsmath}
\usepackage{enumitem}
\usepackage{amssymb,amsmath}
\usepackage{multirow}
\usepackage{mathtools}
\usepackage{bbm}
\usepackage{bm}
\usepackage{amsfonts}
\usepackage{amssymb}
\usepackage{calligra}
\usepackage{calrsfs}
\usepackage{makecell}
\usepackage[normalem]{ulem}
\usepackage[USenglish,american]{babel}

\expandafter\ifx\csname package@font\endcsname\relax\else
 \expandafter\expandafter
 \expandafter\usepackage
 \expandafter\expandafter
 \expandafter{\csname package@font\endcsname}%
\fi
\begin{document}

\title{Quantum gravity phenomenology and the blackbody radiation}

\author{R. Turcati}%
\email{turcati@cbpf.br} 
\thanks{(Corresponding author)}
\affiliation{Centro Brasileiro de Pesquisas Físicas, Rua Dr. Xavier Sigaud, 150, URCA, Rio de Janeiro CEP 22290-180, RJ, Brazil}%

\author{I. Soares}%
\email{winacio@cbpf.br}
\affiliation{Centro Brasileiro de Pesquisas Físicas, Rua Dr. Xavier Sigaud, 150, URCA, Rio de Janeiro CEP 22290-180, RJ, Brazil}%

\author{S. B. Duarte}%
\email{sbd@cbpf.br}
\affiliation{Centro Brasileiro de Pesquisas Físicas, Rua Dr. Xavier Sigaud, 150, URCA, Rio de Janeiro CEP 22290-180, RJ, Brazil}%

\begin{abstract}

We analyze the blackbody radiation problem in the presence of quantum gravity effects encoded in modified dispersion relations. The spectral radiance and the generalized Stefan-Boltzmann law are studied in this context. Furthermore, the regime of low temperatures is also contemplated, where features related to the blackbody thermal laws and the thermodynamic quantities such as energy, pressure, entropy, and specific heat are obtained. Possible implications in compact objects such as neutron stars are also discussed.

\end{abstract}

\maketitle

\section{Introduction}\label{introduction}

The nature of spacetime is perhaps the main challenge in modern physics. 
It is commonly accepted that at the Planck scale, i.e., for energies near $10^{19}GeV$, our physical theories cease to be predictive, and quantum gravity effects should become relevant \cite{Gross:1987ar,Maggiore:1993rv,Addazi:2021xuf}. Nevertheless, such an energy scale is far beyond our technological limitations, which has hampered any direct observation of these relics signatures.

On the other hand, in the past years, it has been suggested that quantum gravity effects may induce small deviations in the standard predictions, which could be observable, at least in principle, through astronomical observations and ground experiments \cite{Amelino-Camelia:1999hpv,Addazi:2021xuf,Das:2008kaa,Ali:2011fa,Bosso:2018ckz,Barausse:2020rsu,LISACosmologyWorkingGroup:2022jok,Bosso:2016ycv}. The possibility of observing these phenomena includes neutrino oscillations, gamma rays, laser interferometry, and CPT violation, among others \cite{Kostelecky:2003fs}. 

The framework where these phenomena can be investigated is the so-called Quantum Gravity Phenomenology\cite{Ali:2009zq}. 
This formalism includes several tests such as prints on the initial cosmological perturbations, black holes related to extra dimensions, and Planck scale spacetime fuzziness. Features related to violations of fundamental symmetries like the Lorentz and CPT symmetries are also contemplated in these scenarios\cite{Kostelecky:2003fs}. 

Among the several ways to probe these signatures arising from quantum gravity, a major role is played through modifications of the dispersion relation\cite{Amelino-Camelia:1997ieq,Ellis:1999rz,Verma:2018lfc,Pal:2019awn}. Deviations from the standard relativistic dispersion relations can be studied by analyzing high-energy photons emitted by astrophysical sources at cosmological distances, which may sign quantum gravity effects or even the presence of Lorentz symmetry violations\cite{Amelino-Camelia:2009imt}. In loop quantum gravity scenarios, for instance, some works have indicated that Lorentz symmetry violations may induce modifications in the Maxwell equations, leading to birefringence effects in photon propagation\cite{Gambini:1998it}.

Modified dispersion relations and higher-order derivatives field theories are common aspects of several phenomenological approaches to quantum gravity. Indeed, in the framework of quantum gravity phenomenology, modifications of the relativistic dispersion relation associated with free particles are expected at the Planck scale, which can shed some light on the underlying quantum theory of gravity. From the phenomenological point of view, two possibilities must be distinguished. The first introduces preferred frames in the spacetime structure, and the Lorentz symmetry is not preserved. This approach is set in the framework of the effective field theory, where several experiments and observations in the past years led to many constraints in the coefficients responsible for the breakdown of the Lorentz invariance\cite{Kostelecky:2008ts}. The second one preserves the equivalence of all inertial observers in the so-called deformed special relativity, where is introduced 
a new observer-independent scale, the Planck energy\cite{Amelino-Camelia:2010lsq}. This assumption, in turn, requires a deformation of the relativistic symmetries of special relativity, which may be obtained by adopting a non-standard representation of the boost generator. 

On the other hand, the Planck scale is expected to suppress the modifications of the relativistic dispersion relations. In this vein, one might wonder if these effects are significant and why consider them after all. From the cosmology perspective, trans-Planckian effects should have affected the early-stage evolution of the universe, which could have led to fingerprints in the cosmic microwave background\cite{Easther:2001fi,Kempf:2001fa}. Furthermore, the thermal emission of black holes is associated with modes of arbitrarily high energies near the horizon, which have been based on modified dispersion relations, and have provided remarkable insights with regards to the Hawking radiation\cite{Jacobson:1991gr}. In addition, Lorentz violation effects in perturbative quantum field theories could be enhanced in the mechanism of regularization and renormalization\cite{Visser:2009fg}. 

Phenomenological studies involving thermodynamic properties of a photon gas with modified dispersion relations have been investigated in previous works\cite{Camacho:2007qy,Zhang:2011ms,Das:2010gk,Chandra:2011nj,Faruk:2016brl,Chung:2018viu,Bosso:2021agx}, whereas thermodynamics aspects of bosons and fermions applied to cosmology and astrophysics have been discussed in similar scenarios\cite{Alexander:2001ck,Bertolami:2009wa}. Thus, there exists a large literature associated with this subject, which shows considerable interest in this topic. Our approach intends to provide details about the blackbody radiation thermal laws and probe the eventual implications in neutron stars.

Here, we will adopt a pragmatic point of view and try to understand the eventual consequences of the leading order correction of a general energy-dependent dispersion relation in the blackbody radiation laws and the related thermodynamic quantities. 

The structure of this paper is organized as follows: In Sec. (\ref{MDR}) modified dispersion relations are introduced and the grand-canonical formalism is employed for these dispersion relations. In Sec. (\ref{deformedPlanck}), the spectral radiance and the Stefan-Boltzmann law in the present context are derived. The blackbody thermal laws and the thermodynamic quantities in the regime of low temperatures are contemplated in Secs. (\ref{lowT}) and (\ref{Thermoquantities}). Our final remarks and further perspectives can be found in Sec. (\ref{conclusions}).



\section{Modified dispersion relation and the statistical mechanics}\label{MDR}

Motivated by the idea that quantum gravity effects may induce modifications in the dispersion relation \cite{Amelino-Camelia:2002siu,Alfaro:2002ya,Amelino-Camelia:2004qyt,Alfaro:2004ur,Magueijo:2004vv,Amelino-Camelia:2005xme}, we consider the general energy-dependent dispersion relation of the following form:
\begin{eqnarray}\label{fullDR}
%
%
c^{2}\mathbf{p}^{2}=f\left(E,m;\lambda_{n}\right)\approx{E}^{2}-m^{2}c^{4}+\sum_{n=1}^{\infty}\lambda_{n}E^{(n+2)},    
\end{eqnarray}
where $f\left(E,m;\lambda_{n}\right)$ is the function that gives the exact dispersion relation arising from some fundamental quantum gravity theory, $\lambda_{n}$ are factors with the dimension of the inverse of energy, where its value depends on the specific quantum gravity model, and we have assumed a Taylor expansion for $E\ll1/{\lambda}_{n}$. In addition, whenever $\lambda_{n}\to0$, one recovers the standard relativistic dispersion relation. 

Since we are considering the gauge sector of the electromagnetism, we will then assume a zero rest mass. Furthermore, we will focus our attention on the first-order correction, which reduces (\ref{fullDR}) to
\begin{eqnarray}\label{MDR}
c^{2}\mathbf{p}^{2}=E^{2}+\lambda_{1}E^{3}, 
\end{eqnarray}
where $\lambda_{1}$ is believed to be of the order of the Planck length $l_{P}$. The modified dispersion relation (\ref{MDR}) is the fundamental equation for the dynamics of the photons from which the blackbody thermodynamic properties will be studied.

At this point, some comments are in order.
\begin{enumerate}
    \item For $\lambda_{1}$ positive-definite, the momentum $p$ is always positive.

    \item $\lambda_{1}$ negative and $|\lambda_{1}|<1/E$ provides a physical branch for the particle.

    \item Whenever $\lambda_{1}>1/E$ for negative values of $\lambda_{1}$, it corresponds to a nonphysical region. 
\end{enumerate}

Since the quantum gravity effects are introduced through modifications of the relativistic dispersion relations, one could ask what the consequences are to the statistical mechanics systems. In this sense, it is important to stress the fact that the foundations of statistical mechanics are preserved in this situation\cite{Alexander:2001ck,Landau:1980mil}. However, there will be modifications in the density of accessible states, which will be discussed later. 

At this point, it would be interesting to ask ourselves about the reason for doing research with modified dispersion relations of the form (\ref{fullDR}) in the photon sector. The increased interest in recent years in this subject is mainly motivated by the possibility of detecting the energy-dependent time of arrival of gamma-ray bursts originating at cosmological distances\cite{Amelino-Camelia:1997ieq,Ellis:1999rz}. In this context, modified dispersion relations with higher order terms in the energy lead to an energy-dependent speed of massless particles, given by $v\approx1-\lambda{E}$. A non-constant speed of light in the vacuum, in turn, implies a retardation time (positive $\lambda$) for a signal propagating with high energy, characterizing a subliminal propagation for the photons. From the theoretical perspective, the energy-dependent speed of light may solve the horizon, flatness, and cosmological constant problems\cite{Albrecht:1998ir}. In addition, some results in loop quantum gravity suggest that modified dispersion relations of the form (\ref{fullDR}) in the photon and neutrinos sectors may emerge in connection to the Lorentz symmetry violation\cite{Li:2022szn}, which provides a consistent framework to explore the phenomenological consequences of quantum gravity theories. Furthermore, modified dispersion relations emerge in theories where the uncertainty principle is generalized, a common aspect of theories with a minimal length\cite{Hossenfelder:2005ed,Kempf:1994su}. Although some theoretical frameworks motivate modifications in the relativistic dispersion relations, the fact that modified dispersion relations are experimentally measurable by themselves is, therefore, a very compelling reason to consider this seriously on their own. 

Having motivated the physical relevance of the relation (\ref{fullDR}), let us then see how the density of states for the energy of a photon gas is modified in the present scenario. To begin with, we use the grand-canonical ensemble to describe the photon gas in a container of volume V\cite{Pathria:1996hda}. For a system of $N$ massless bosons with energy spectrum $\beta\hbar{w}_{i}$, we have that each state is labeled by $i$ $\left(i=1,2,...\right)$, where $\beta=1/k_{B}T$, $k_{B}$ is the Boltzmann constant and $T$ is the temperature. In addition, we assume that the De Broglie relation is not modified, leading to boundary conditions of the photons with momenta $p_{n}=n/2L$\cite{Amelino-Camelia:2005zpp,Nozari:2006yia}. 

The connection between thermodynamics and statistical mechanics is given by the gran-canonical potential, 
\begin{eqnarray}
\phi=-\frac{1}{\beta}log\mathcal{Z},    
\end{eqnarray}
where the logarithmic of the partition function takes the following form:
\begin{eqnarray}
log\mathcal{Z}=-\sum_{i}log\left(1-e^{-\beta\hbar{w}_{i}}\right).    
\end{eqnarray}

The average energy is then given by
\begin{eqnarray}
\langle{E}\rangle=\frac{1}{V}\sum_{i}\frac{\hbar{w}_{i}}{e^{\beta\hbar{w_{i}}}-1}.    
\end{eqnarray}

In the large-volume limit, one can approximate the sum by an integral in the following way:
\begin{eqnarray}
\sum_{i}\rightarrow\int{d{\mathbf{x}}\int\frac{d{\mathbf{k}}}{(2\pi)^{3}}}=\frac{4\pi{V}}{(2\pi)^{3}}\int_{0}^{\infty}dkk^{2}.
\end{eqnarray}

Performing an integral variable change from the momenta $k-$space to the frequency $w-$space, in a similar way as it is obtained in statistical mechanics\cite{huang2008statistical}, through the use of the dispersion relation (\ref{MDR}), one obtains
\begin{eqnarray}\label{largevolume}
\sum_{i}\rightarrow\frac{V}{\pi^{2}c^{3}}\int_{0}^{\infty}dww^{2}g\left(\lambda_{1};w\right),    
\end{eqnarray}
where the function $g\left(\lambda_{1};w\right)$ holds the quantum gravity corrections and is given by
\begin{eqnarray}\label{gmodification}
g\left(\lambda_{1};w\right)=\left(1+\lambda_{1}\hbar{w}\right)^{3/2}+\left(\frac{\lambda_{1}\hbar}{2}\right)w\sqrt{1+\lambda_{1}\hbar{w}}.    
\end{eqnarray}

From the above relation, one promptly notes that the density of accessible states has changed from the standard result for photons. It is worth stressing that the expression (\ref{gmodification}) encodes the quantum gravity effects for the photon gas. Whenever $\lambda_{1}=0$, $g=1$, and the usual predictions of the statistical mechanics are recovered. For positive values of $\lambda_{1}$, the $g-$function is always positive-definite, while for negative ones, the constraint $|\lambda|<\hbar{w}$ must be satisfied.

According to the standard prescription, the number $N$ of modes in the momenta interval $\left[k,k+dk\right]$ is
\begin{eqnarray}\label{MDR-3}
g\left(k\right)dk=\frac{V}{2\pi^{2}}k^{2}dk.    
\end{eqnarray}

By performing a straightforward computation, and taking the modified dispersion relation (\ref{MDR}) into account, we promptly find in the frequency $w-$space,
\begin{eqnarray}\label{modifieddensitystate}
%
%
%
g\left(w\right)dw=\frac{Vw^{2}}{\pi^{2}c^{3}}g\left(\lambda_{1};w\right).
\end{eqnarray}

The modified dispersion relation (\ref{MDR-3}) depends only on the magnitude of the momentum vector, i.e., $|\mathbf{p}|=p=\hbar{k}$. Therefore, the integration in the phase space is trivial and yields $V$. In addition, the density of momentum states remains the same since the periodic conditions are not changed. From relation (\ref{largevolume}), becomes clear that the dispersion relation (\ref{MDR-3}) modifies the density of accessible states, which can be seen in relation (\ref{modifieddensitystate}). As a consequence, there will be an increase or decrease in the number of photons for each frequency mode depending on the sign of $\lambda_{1}$, which will influence the behavior of the thermodynamics quantities such as the energy, pressure, entropy, and specific heat.

The classical equipartition theorem, in turn, states that the average energy for photons in thermal equilibrium at temperature T is $k_{B}T$. Quantum mechanics, on the other hand, assert that each mode of a photon gas satisfying the Bose-Einstein statistics has an average energy given by $\bar{E}=\hbar{w}/(e^{\beta\hbar{w}}-1)$. In the present framework, quantum gravity effects modify the average energy, which generalizes to
\begin{eqnarray}
\bar{E}=\left(\frac{\hbar{w}}{e^{\beta\hbar{w}}-1}\right)g\left(\lambda_{1};w\right).    
\end{eqnarray}

\section{The deformed Planck spectrum}\label{deformedPlanck}

For a gas of noninteracting photons in this situation, the energy density in thermal equilibrium at temperature $T$ in the frequency interval $\left[w,w+dw\right]$ is given by
\begin{eqnarray}\label{fullPlanckdeformed}
u\left(w\right)dw&=&\frac{w^{2}}{\pi^{2}c^{3}}\left[\left(1+\lambda_{1}\hbar{w}\right)^{3/2}+\left(\frac{\lambda_{1}\hbar}{2}\right)w\sqrt{1+\lambda_{1}\hbar{w}}\right]\nonumber\\
&&\times\left(\frac{\hbar{w}}{e^{\beta\hbar{w}}-1}\right)dw.
\end{eqnarray}

\begin{center}
\begin{figure}[htb]
\begin{minipage}{0.5\textwidth}
\begin{tikzpicture}
  \node (img)  {\includegraphics[scale=0.65]{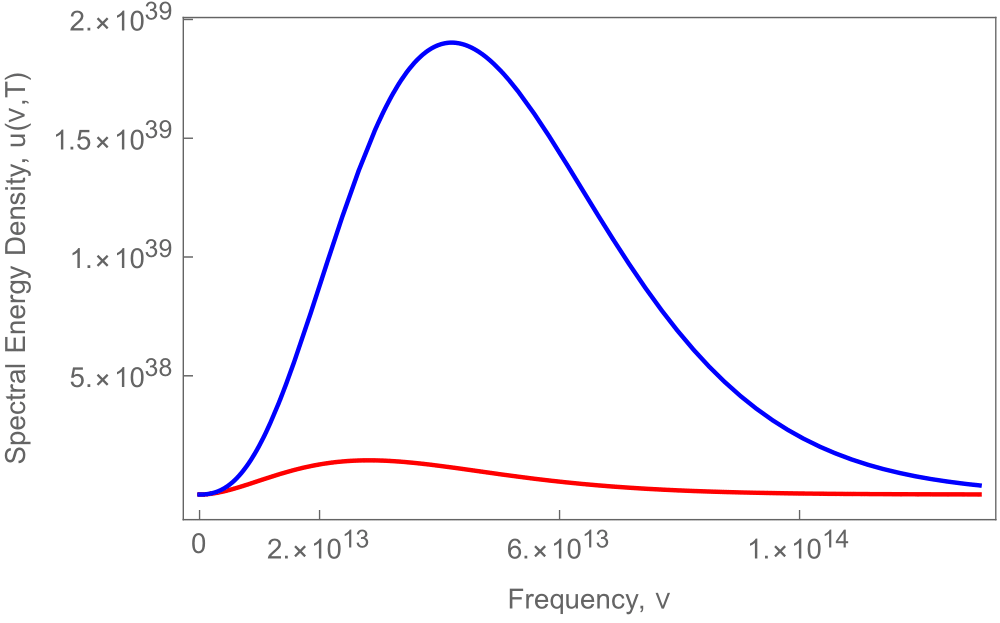}};
 \end{tikzpicture}
\end{minipage}%
\caption{Graph of the spectral energy density distribution for the quantum gravity model under consideration for $T=10^{13}GeV$ and $\lambda_{1}=10^{-13}GeV$. Here, we adopted $c=\hbar=k_{B}=1$. The red line corresponds to the Planck spectrum, while the blue one is associated with the quantum gravity model.}\label{figure1}
\end{figure}
\end{center}

Note that whenever $\lambda_{1}\rightarrow0$, one recovers the Planck distribution. To get a better idea of the changes induced by the quantum gravity effects, we plot, in Fig. (\ref{figure1}), both the Planck distribution and the deformed spectrum. To perform such analysis, we have considered $T\approx1/\lambda_{1}$. There we can see that the quantum gravity effects are relevant when we deal with high temperatures in the Universe, such as the inflationary epoch temperature, for instance, where the temperature reaches $T=10^{13}GeV$.

Usually, one would expect that quantum gravity corrections at the standard thermodynamics results would be neglected at low temperatures. On the other hand, these effects will depend on the value of $\lambda_{1}$. To evaluate if low-temperature effects could be probed at this regime, let us consider the Stefan-Boltzmann law for some values of $\lambda_{1}$.

\begin{center}
\begin{figure}[htb]
\begin{minipage}{0.5\textwidth}
\begin{tikzpicture}
  \node (img)  {\includegraphics[scale=0.65]{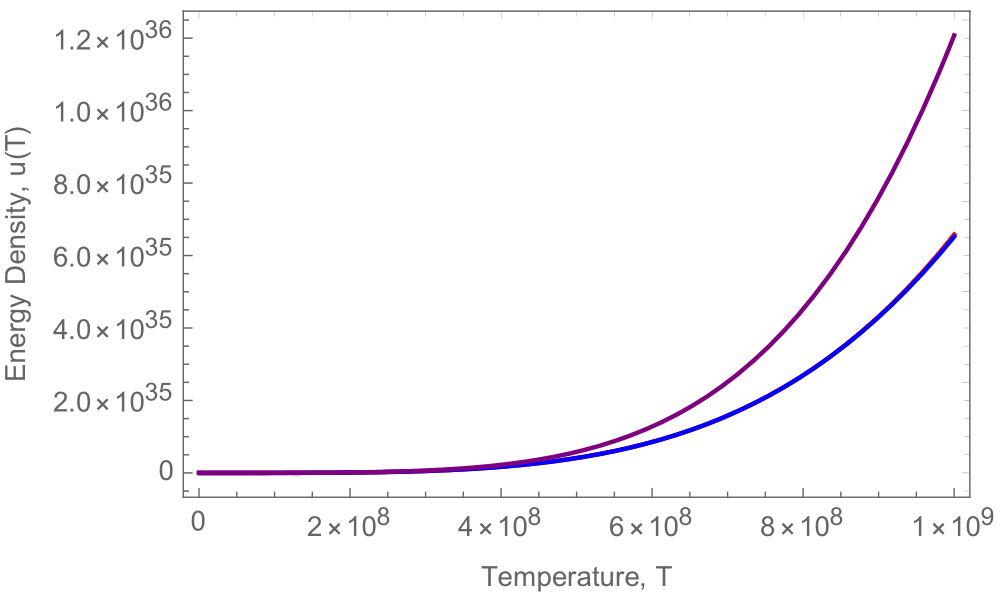}};
 \end{tikzpicture}
\end{minipage}%
\caption{Graph of the energy density vs temperature. Here, we have adopted $\lambda_{1}=10^{-10}eV^{-1}$ (purple line) and $\lambda_{1}=10^{-13}eV^{-1}$ (blue line) for the quantum gravity model under consideration, where the curves were obtained through numerical integration. The standard Stefan-Boltzman law is associated with the red line, which is overlapped by the blue line. To evaluate the above graph, we have assumed natural units.
}\label{figure3}
\end{figure}
\end{center}

Therefore, taking into account the Stefan-Boltzmann law
\begin{eqnarray}
u=\frac{4\sigma}{c}{T}^{4},    
\end{eqnarray}
where 
\begin{eqnarray}\label{SBconstant}
\sigma=\frac{2\pi^{5}k_{B}^{4}}{15h^{3}c^{2}}     
\end{eqnarray}
is the Stefan-Boltzmann constant, we obtain the following curves in Fig (\ref{figure3}). For $\lambda_{1}=10^{-10}eV^{-1}$ (purple line), one promptly notes that for temperatures near $10^{8}eV$, small deviations from the standard Stefan-Boltzman law can be observed. For values of $\lambda_{1}$ around $10^{-10}eV^{-1}$, we can study the low-temperature regime and investigate how the thermodynamics of the blackbody radiation is affected.

\section{Quantum gravity corrections at low temperature}\label{lowT}

Now, we would like to identify the leading quantum gravity correction to the thermodynamic quantities at low temperatures, i.e., when $\beta\ll1$. In this regime, the factor $(e^{\beta\hbar{w}}-1)^{-1}$ decays exponentially, and then one can expand $g(\lambda_{1},w)$ in Taylor series around zero. To perform this approximation, the condition 
$\lambda_{1}\ll\beta$ must be satisfied. For compact objects such as newly formed neutron stars, which hold temperatures around $10^{11}-10^{12}K$ inside its interior, this condition is ensured if $\lambda_{1}\ll10GeV^{-1}$. 

Taking into account the mentioned assumptions, therefore, Eq. (\ref{fullPlanckdeformed}) reduces to
\begin{eqnarray}\label{lowregimePlanck}
u\left(w\right)dw=\frac{w^{2}}{\pi^{2}c^{3}}\left(1+2\lambda_{1}\hbar{w}\right)\left(\frac{\hbar{w}}{e^{\beta\hbar{w}}-1}\right)dw.    
\end{eqnarray}

\begin{center}
\begin{figure}[htb]
\begin{minipage}{0.5\textwidth}
\begin{tikzpicture}
  \node (img)  {\includegraphics[scale=0.65]{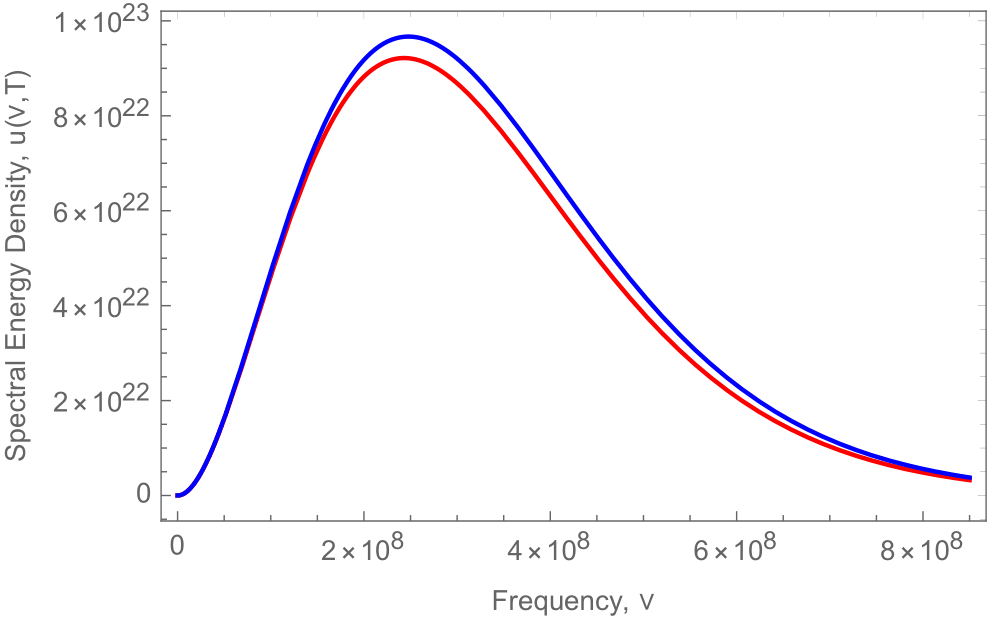}};
 \end{tikzpicture}
\end{minipage}%
\caption{Graph of the spectral energy density distribution for the quantum gravity model under consideration for $T=86,17MeV$ and $\lambda_{1}=10GeV^{-1}$. Here, we adopted $c=\hbar=k_{B}=1$. The red line corresponds to the Planck spectrum, while the blue one is associated with the quantum gravity model.}\label{figure2}
\end{figure}
\end{center}

The maximum value of the above relation, as a function of $w$, is shifted, modifying Wien's displacement law (See Fig. \ref{figure2}). In the regime of long wavelength, the above frequency distribution takes the following form:
\begin{eqnarray}
u\left(w\right)=\frac{w^{2}}{\pi^{2}c^{3}}\left(1+2\lambda_{1}\hbar{w}\right)k_{B}T,    
\end{eqnarray}
which provides us with the quantum gravity corrections to the Rayleigh-Jeans formula.

The total energy, in turn, can now be computed. To accomplish that, one must integrate Eq. (\ref{lowregimePlanck}), where, taking into account the integral\cite{Abramowitz1974}
\begin{eqnarray}
\int^{\infty}_{0}dx\frac{x^{n}}{e^{x}-1}=\Gamma\left(n+1\right)\zeta\left(n+1\right),    
\end{eqnarray}
will give us
\begin{eqnarray}\label{energyLowT}
u\left(T\right)=\frac{\pi^{2}k_{B}^{4}T^{4}}{15c^{3}\hbar^{3}}+\frac{48\lambda_{1}\xi\left(5\right)k_{B}^{5}T^{5}}{\pi^{2}c^{3}\hbar^{3}},
\end{eqnarray}
with $\Gamma\left(n+1\right)$ being the Gamma function and $\zeta\left(n+1\right)$ being the Riemann zeta function.

If one introduces the Stefan-Boltzmann constant (\ref{SBconstant}) 
in the above expression, one promptly gets
\begin{eqnarray}\label{u2}
u\left(T\right)=\frac{4\sigma}{c}{T^{4}}+\frac{48\lambda_{1}\xi\left(5\right)k_{B}^{5}T^{5}}{\pi^{2}c^{3}\hbar^{3}}.
\end{eqnarray}

For $\lambda_{1}\rightarrow0$, one recovers the Stefan-Boltzmann law. Equation (\ref{u2}) can also be written as
\begin{eqnarray}\label{u2}
u\left(T\right)=\frac{4}{c}\sigma_{eff}\left(\lambda_{1};T\right){T^{4}},
\end{eqnarray}
where
\begin{eqnarray}\label{u2}
\sigma_{eff}\left(\lambda_{1};T\right)=\sigma\left(1+\frac{720\lambda_{1}\xi\left(5\right)}{\pi^{4}}k_{B}T\right)
\end{eqnarray}
is the effective Stefan-Boltzmann constant.

For high temperatures in dense stars, for instance, it is possible to observe deviations from the Stefan-Boltzmann law. 

\subsection{Thermodynamics in the presence of modified dispersion relations}\label{Thermoquantities}

One of the most important results of modifying the relativistic dispersion relation is the deformation of thermodynamical quantities. To investigate this, we take into account the grand canonical potential, which can be used to derive the pressure, entropy, and specific heat for the blackbody radiation. 

For the pressure, we then find
\begin{eqnarray}\label{pressureLowT}
p=\frac{\pi^{2}k_{B}^{4}T^{4}}{45c^{3}\hbar^{3}}+\frac{12\lambda_{1}\xi\left(5\right)k_{B}^{5}T^{5}}{\pi^{2}c^{3}\hbar^{3}},
\end{eqnarray}
while for the entropy and heat capacity densities, we have
\begin{eqnarray}
s=\frac{4\pi^{2}k_{B}^{4}T^{3}}{45c^{3}\hbar^{3}}+\frac{60\lambda_{1}\xi\left(5\right)k_{B}^{5}T^{4}}{\pi^{2}c^{3}\hbar^{3}},
\end{eqnarray}
and
\begin{eqnarray}\label{specificheat}
c_{V}=\frac{4\pi^{2}k_{B}^{4}T^{3}}{15c^{3}\hbar^{3}}+\frac{240\lambda_{1}\xi\left(5\right)k_{B}^{5}T^{4}}{\pi^{2}c^{3}\hbar^{3}}.
\end{eqnarray}

We would like to stress that depending on the sign of $\lambda_{1}$, the above thermodynamic quantities can increase or decrease, which can influence the dynamics of compact objects such as neutron stars. In addition, for temperatures that are high enough, departures from the standard results of thermodynamics should be measurable for these objects.

Let us now analyze the case for $\lambda_{1}$ positive values. In this scenario, the pressure grows concerning the ideal gas. 
This result suggests that the quantum gravity effects may be measured as a repulsive interaction among the components of the photon gas. This interpretation can be enlightened if one considers the deviations from the ideal gas in terms of virial expansion\cite{huang2008statistical}, where the pressure takes the following form:
\begin{eqnarray}
\frac{pV}{Nk_{B}T}=\sum_{l=1}^{\infty}B_{l}\left(T\right)\left(\frac{N\lambda^{3}}{V}\right)^{l-1},    
\end{eqnarray}
with $\lambda=\sqrt{2\pi\hbar^{2}/mk_{B}T}$ being the thermal wavelength, $N/V$ denotes the particle density, $V$ is the volume and $B_{l}(T)$ is the virial coefficient. For $l=1$, $B_{1}=1$, and we recover the equation of state for an ideal gas. For $l=2$, the second virial coefficient, $B_{2}$, is a generic function of the pair potential between the particles of the gas. A repulsive interaction means $B_{2}(T)>0$, which increases the pressure, while $B_{2}(T)<0$ implies an attractive interaction, leading to a lower pressure than the ideal gas.

The preceding analysis indicates that the quantum gravity leads to deviations in the pressure for the ideal gas. For $\lambda_{1}$ positive, the pressure grows, while for negative values of $\lambda_{1}$, the pressure decreases.

The specific heat (\ref{specificheat}) is another measurable quantity that can be used for the search of quantum gravity corrections at the level of experimental setups. However, these modifications appear as $\lambda_{1}T$, which introduces rigid bounds from the experimental point of view.

The equation of state, in turn, can be found by dividing the pressure (\ref{pressureLowT}) by the energy (\ref{energyLowT}), which gives us
\begin{eqnarray}\label{eos}
\frac{p}{u}=\frac{1}{3}-\frac{1620\lambda_{1}\xi(5)k_{B}T}{\pi^{4}}.   
\end{eqnarray}

In the limit $\lambda_{1}\rightarrow0$, we have $p=3u$. The equation of state is a function of the energy, i.e., $p=p(u)$. To eliminate the temperature dependence in the above expression, one needs to expand the temperature in power series of $\lambda$, up to the first order, as $T=T_{0}+\lambda_{1}T_{1}$ and apply it recursively in the energy density (\ref{u2}). Therefore, comparing the powers in $\lambda_{1}$, one promptly finds for the temperature the following form:
\begin{eqnarray}\label{temperature}
T=\left(\frac{15\hbar^{3}c^{3}}{\pi^{2}}\right)^{1/4}\frac{u^{1/4}}{k_{B}}+\frac{1620\xi(5)}{\pi^{5}}\left(15\hbar^{3}c^{3}\right)^{1/2}\frac{\lambda_{1}}{k_{B}}u^{1/2}.\nonumber\\    
\end{eqnarray}

Inserting relation (\ref{temperature}) in (\ref{eos}), we have, up to the linear order in $\lambda_{1}$, 
\begin{eqnarray}\label{eosfinal}
p=\frac{u}{3}\left[1-\frac{540\lambda_{1}\xi(5)}{\pi^{4}}\left(\frac{15\hbar^{3}c^{3}}{\pi^{2}}\right)^{1/4}\lambda_{1}u^{1/4}\right]. 
\end{eqnarray}

From the above equation, one promptly notes the emergence of quantum gravity effects in the standard equation of state of the order $\lambda_{1}(\hbar{c})^{3/4}u^{1/4}$. 

\section{Final Remarks}\label{conclusions}

In this work, we have considered the effect of quantum gravity effects in the thermodynamics of blackbody radiation. Starting from an energy-dependent dispersion relation up to the leading order correction, we have explored the behavior of the emission spectrum in both high and low temperatures. With regards to the high-temperature regime, the deformed Planck distribution and the generalized Stefan-Boltzmann law were derived. In the low-temperature limit, on the other hand, the blackbody radiation laws and the thermodynamics properties were obtained. In this regime, we have shown that for positive values of $\lambda_{1}$, the energy density, the pressure, the entropy, and the specific heat are larger than those of special relativity. In contrast, with a negative $\lambda_{1}$, the quantum gravity effects emerge in the opposite direction. Nevertheless, these modifications manifest themselves significantly when the temperature is larger than $T=10^{12}K$, which cannot be directly detected in lab setups. On the other hand, it suggests that these predictions may play an important role in the early universe, such as inflation, and in the formation of neutron stars.

Regarding the value of $\lambda_{1}$, there is no experimental evidence of deviations from the relativistic dispersion relation up to the TeV energy scale\cite{ParticleDataGroup:2008zun}. Therefore, values of $\lambda_{1}\ll10^{-3}GeV^{-1}$ are a suitable choice for considering eventual quantum gravity effects in the photonic sector at low energies in our framework. Furthermore, recent measurements of gamma-ray bursts suggest strong constraints on the in-vacuo dispersion relation of these high-energetic photons. Different approaches to quantum gravity lead to similar results when taking into account the dispersion relation of the form (\ref{MDR}), with the speed of light taking the following form:
\begin{eqnarray}
v\left(E\right)\approx1-\frac{E}{E_{QG}},   
\end{eqnarray}
where $E_{QG}\approx3.6\times10^{17}GeV$\cite{Amelino-Camelia:2017zva}, which provide an estimative value around $\lambda_{1}<2.7\times10^{-18}(GeV)^{-1}$. These estimates justify the consideration of small values for $\lambda_{1}$, which gives support to the idea of considering the effects of the leading order of the dispersion relation of the form (\ref{fullDR}).

At very high temperatures, the blackbody radiation thermal laws induced by the low-energy expansion (\ref{fullDR}) are not valid anymore. We have thus provided a reliable picture at temperatures well below the Planck scale. A full-order modified dispersion relation for the photonic sector of the form $c^{2}p^{2}=E^{2}\left[1+\mathcal{F}\left(E/E_{QG}\right)\right]$, with $E_{QG}$ being the quantum gravity energy scale and $\mathcal{F}$ being a generic model-dependent function, is expected to emerge for temperatures $T/E_{QG}\gg1$, which is supposed to eliminate such divergences. 

We also would like to draw the reader's attention to the {\it soccer ball problem}\cite{Hossenfelder:2014ifa}. In this work, we have started with the partition function in the grand canonical ensemble and then applied the standard methodology of statistical mechanics to derive the corresponding blackbody thermal laws. Theories in which there are nonlinear modifications of composition law momenta, such as deformed special relativity, lead to conceptual problems associated with the transformation of multi-particle states. In particular, with successive addition of many particles with sub-Planck energies each, one may end up with a huge energy contribution greater than the Planck energy when considering a macroscopic object. In our approach, the deformed Planck distribution (\ref{fullPlanckdeformed}), which is valid for temperatures below the Planck scale, is well-behaved over the frequency spectrum (see Fig. (\ref{figure1})), which suggests that the mentioned problem does not show up in our framework. For very high temperatures, as commented above, it is expected a full-order modified dispersion relation, where the features related to the transformation of the multiparticle states may lead to other possibilities. This analysis is out of the scope of this paper.

Some other theories have considered higher-order terms in relation (\ref{fullDR}). In Ref. \cite{Nozari:2006yia}, based on arguments of black hole thermodynamics, only even powers of the energy in the modified dispersion relation were considered. Here, on the other hand, we have considered the leading order term in the mentioned expansion. Higher-order terms ($n\ge2$) would lead to temperature dependences of the form $\lambda_{n}T^{4+n}$ to the Stefan-Boltzmann law. Furthermore, the profile of the blackbody spectrum would also be modified by these terms.

As a prospect, we intend to extend our formalism here developed to massive particles in the fermionic sector and, then, study the consequences of the hydrostatic equilibrium of compact objects such as neutron stars. The massive bosonic sector can also be contemplated. Since the density of accessible states for massless bosonic particles is modified in the present scenario, one would expect the same behavior for massive particles. In this sense, it seems to suggest that quantum gravity corrections should play a role in the Bose-Einstein condensate phenomena, leading to a modification at the condensation temperature, which could take to measurable experiments in the foreseeable future\cite{Das:2021skl}.

\section{Acknowledgements}

This work is a part of the INCT-FNA project proc. No. 464898/2014-5. RT acknowledges financial support from the PCI program of the Brazilian agency Conselho Nacional de Desenvolvimento Científico e Tecnológico – CNPq. SBD thanks CNPq for partial financial support.

\bibliographystyle{apsrev4-1}
\bibliography{Bibliography.bib}

\end{document}